\documentclass[superscriptaddress,showpacs,twocolumn,prl]{revtex4}

\pdfoutput=1 
\usepackage{graphicx,epstopdf,float}
\usepackage{bbm}
\usepackage{bm}
\usepackage{mathbbol}
\usepackage{amsmath}
\usepackage{longtable}
\usepackage{amssymb}
\usepackage{bbold}
\usepackage{color}
\usepackage{braket}
\usepackage{soul}

\newcommand{\beq}{\begin{equation}}
\newcommand{\eeq}{\end{equation}}

\definecolor{Red}{rgb}{1,0,0}
\definecolor{Blue}{rgb}{0,0,1}

\begin{document}
\title{Optimal interfacing a GHz-bandwidth heralded single photon source with on-demand, broadband quantum memories}

\author{P.~S.~Michelberger}
\affiliation{Clarendon Laboratory, University of Oxford, Parks Road, Oxford OX1 3PU, UK}

\author{M.~Karpi\'nski}
\affiliation{Clarendon Laboratory, University of Oxford, Parks Road, Oxford OX1 3PU, UK}

\author{I.~A.~Walmsley}
\affiliation{Clarendon Laboratory, University of Oxford, Parks Road, Oxford OX1 3PU, UK}

\author{J.~Nunn}
\affiliation{Clarendon Laboratory, University of Oxford, Parks Road, Oxford OX1 3PU, UK}

%
%\date{\today}

%
\begin{abstract}
Photonics offers a route to fast and distributed quantum computing in ambient conditions, provided that photon sources and logic gates can be operated deterministically. Quantum memories, capable of storing and re-emitting photons on demand, enable quasi-deterministic operations by synchronising stochastic events. We recently interfaced a Raman-type quantum memory with a traveling-wave heralded photon source \cite{Michelberger:2014}. Here we discuss the tradespace for the spectral characteristics of such sources, and we present measurements of our source, which represents a practical compromise enabling passive stability, high brightness in a single-pass configuration, high purity and good matching to our Cs vapour memory.
\end{abstract}

\maketitle

%%%%% 				INTRODUCTION 				       	%%%%%
\section{Introduction\label{intro}}
Optical qubits have been the base for many primitive demonstrations of quantum information protocols
thanks to the ease in their manipulation and their low noise floor in room-temperature environments \cite{Knill:2001nx,OBrien:2003fk,Ursin:2007vn,OBrien:2007zr,Spring:2013aa}. 
Unfortunately linear-optical logic gates are non-deterministic, as they rely on particular measurement outcomes to herald the generation of entanglement. The success probability of concatenated non-deterministic gates falls exponentially, and this has prevented the development of large scale quantum photonic information processors. The same scaling problem applies to parametric single-photon sources such as heralded downconversion. They operate at room temperature and produce high quality single photons with high brightness, but it has not been possible to simultaneously run many 
sources in parallel, as the combined success probability becomes too low.

A promising route to scalable quantum photonics, which retains the capability to operate at room temperature, in ambient conditions, with high bandwidths, is to develop mutually compatible downconversion sources and quantum memories, so that photons can be stored and synchronised. Non-deterministic sources and gates can then be multiplexed to achieve quasi-deterministic operation \cite{Nunn:2013ab}. Running such a system at a clock-rate in the GHz regime is ideal as this exploits the fastest speeds available for electronic processing, avoids significant optical dispersion, and is robust to MHz-scale frequency fluctuations in free-running lasers.

\begin{figure}[t]
\centering
\begin{minipage}[l]{0.48\textwidth}
\centering
\includegraphics[width= \textwidth]{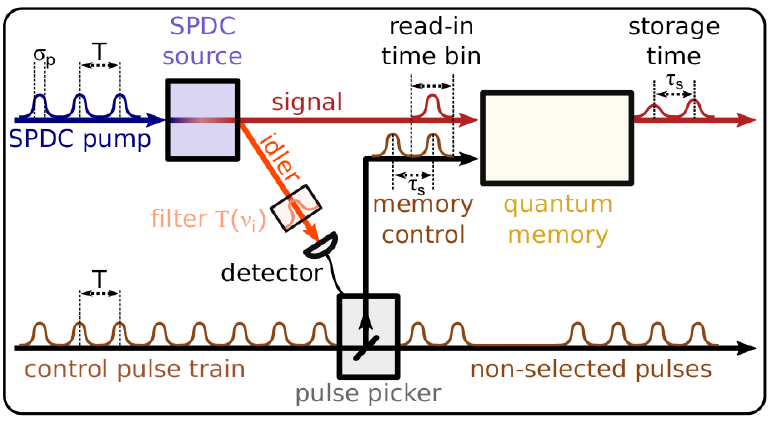}
\end{minipage}
\caption{Schematic of a travelling-wave SPDC source with a quantum memory. On demand storage of heralded single photons is triggered by idler detection events, upon which a memory control pulse is selected from a pulse train for simultaneous insertion into the memory medium. Application of a second control pulse triggers the release of the stored single photon from the memory.}
\label{fig_1}
\end{figure}

We have developed a GHz bandwidth optical memory based on stimulated Raman absorption in warm Cs vapour, and we recently interfaced this memory with a traveling-wave heralded downconversion single photon source \cite{Michelberger:2014}, illustrated in Figure~\ref{fig_1}. 
In this paper we explore the inherent trade-off between purity and brightness in such sources, which arises from time gating introduced by the memory's storage and retrieval operation, and 
report measurements of the spectral characteristics of this source.
We identify a continuous range of optimal configurations for pumping and heralding bandwidths, the two design determinants: practical considerations then dictate how to choose these.

\section{A generic model for a GHz-bandwidth parametric source\label{sec_1}}

\paragraph{Spectral output of heralded down-conversion}

\begin{figure*}[t]
\centering
\includegraphics[width=0.9\textwidth]{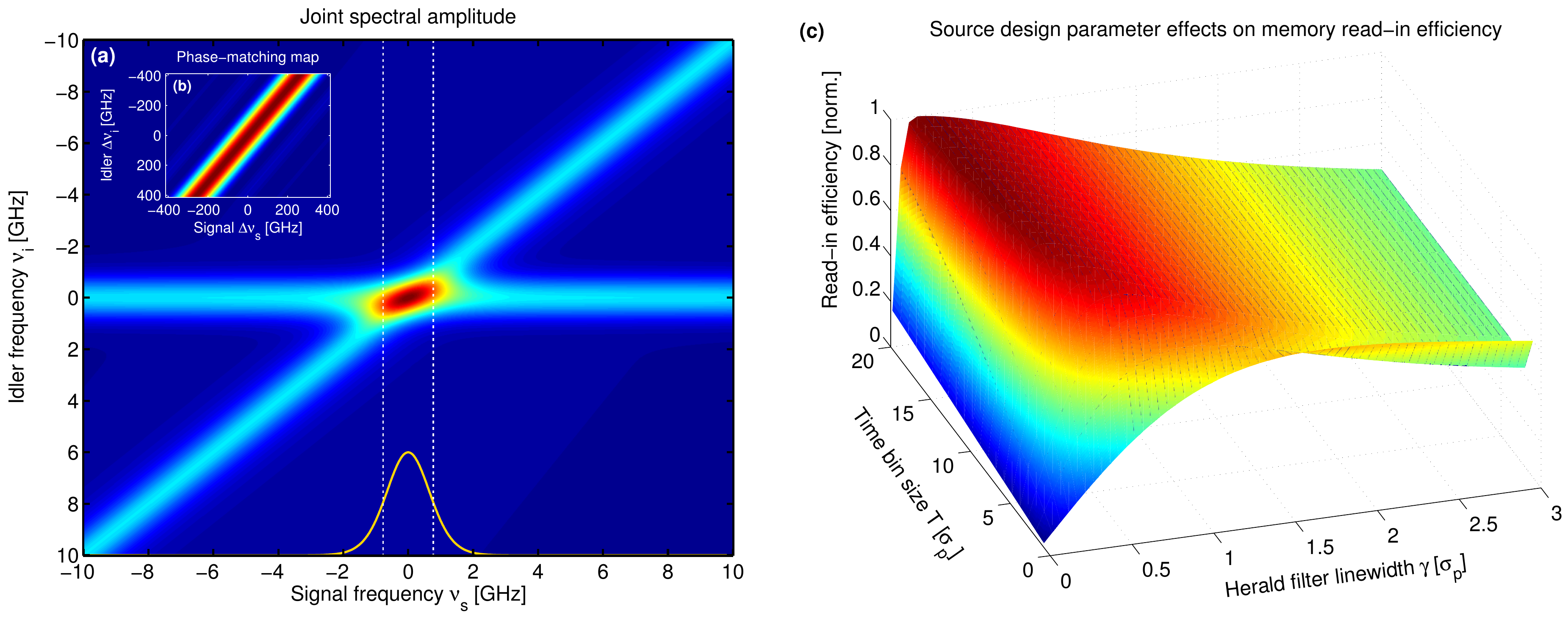}
\caption{(a): Constituents and resulting filtered JSA for GHz-bandwidth SPDC in $\nu_\text{s}$ - $\nu_\text{i}$ space. The diagonal and horizontal stripes illustrate the frequency maps for the SPDC pump and the idler filter, respectively. The former determines the JSA of the unfiltered SPDC output, due to the SPDC's large phase-matching bandwidth (shown in inset (b) \cite{SPDC:note}). The yellow line along the $\nu_\text{s}$ - axis, displays the marginal spectrum of single photons, obtained from heralding by idler detection; dotted vertical lines mark its FWHM bandwidth. 
(c) Influence of SPDC source repetition rate $T$ and idler filter bandwidth $\gamma$ (both in units of SPDC pump bandwidth $\sigma_\text{p}$) on the memory read-in efficiency $\eta_\text{in}$ of the produced heralded single photons, compared to a perfectly mode matched input. 
}
\label{fig_2}
\end{figure*}

Parametric single-photon sources, based on downconversion in a non-linear medium, rely on heralding the presence of a signal photon by detection of a correlated idler photon. A strong pump pulse $\Omega(t)$ drives a multi-mode pairwise squeezing interaction that generates signal and idler photons in pairs, producing the bi-photon state
\begin{equation}
\ket{\psi} = \ket{00} + \int \!\!\!\int \widetilde{f}(\omega_\mathrm{s},\omega_\mathrm{i}) \hat{a}_\mathrm{s}^\dagger(\omega_\mathrm{s}) \hat{a}_\mathrm{i}^\dagger(\omega_\mathrm{i})\,\mathrm{d}\omega_\mathrm{s}\mathrm{d}\omega_\mathrm{i}\ket{00},
\label{eq1_spdc_state}
\end{equation}
where the signal and idler annihilation operators $\hat{a}_\mathrm{s,i}(\omega_\mathrm{s,i})$ destroy photons at frequencies $\omega_\mathrm{s,i}$, respectively, and where the joint spectral amplitude (JSA) 
$\widetilde{f}(\omega_\mathrm{s},\omega_\mathrm{i}) = \tilde{ \Omega }(\omega_\text{s} + \omega_\text{i}) \cdot \tilde{\Phi} (\omega_\text{s},\omega_\text{i})$ 
describes the spectral correlations between the signal and idler modes. 
For traveling-wave sources, schematised in Figure~\ref{fig_1}, the joint spectrum is determined by momentum and energy conservation, entering Equation~\ref{eq1_spdc_state} via the phase-matching function $\tilde{\Phi} (\omega_\text{s},\omega_\text{i})$ and the pump envelope $\tilde{ \Omega} (\omega_\text{s} + \omega_\text{i})$, respectively. 
Figure~\ref{fig_2}~(a) and (b) illustrate both constituents, as well as the resulting JSA, for a GHz-bandwidth source.  
Here, the variations in the material dispersion over the joint spectrum can be neglected, so
the joint spectrum takes the simpler form
%$$
\begin{equation}
\widetilde{f}(\omega_\mathrm{s},\omega_\mathrm{i}) = \widetilde{\Omega}(\omega_\mathrm{s} + \omega_\mathrm{i}). 
\end{equation}
The JSA describes strong frequency anti-correlations between the signal and idler modes, such that detection of an idler photon (without frequency resolution) heralds the presence of a signal photon over a range of possible frequencies: a spectrally mixed state that cannot be directly used for quantum information applications. The spectral purity of the heralded signal photons can be improved by passing the idler photons through a spectral filter with amplitude transmission $\widetilde{\mathcal{T}}(\omega_\mathrm{i})$, which results in a modified joint spectrum
%$$
\begin{equation}
\widetilde{f}_\mathcal{T}(\omega_\mathrm{s},\omega_\mathrm{i}) = \widetilde{\mathcal{T}}(\omega_\mathrm{i})\cdot \widetilde{\Omega}(\omega_\mathrm{s} + \omega_\mathrm{i}),
\end{equation}
%$$
exemplified in Figure~\ref{fig_2}~(a).
Filtering cuts down the stripe-shaped JSA to an ellipse in $\omega_\text{s}$-$\omega_\text{i}$ space. 
However, unless the ellipses' symmetry axes align with the $\omega_\text{s}$-$\omega_\text{i}$ coordinate axes, heralding by idler photon detection will still produce a multi-mode state. 
We can see this, when expressing the filtered bi-photon state in terms of its Schmidt decomposition \cite{Eberly:2006}, using broadband mode operators \cite{Moseley:2008}. 
Its density matrix $\hat{\rho} = \ket{\psi} \bra{\psi}$ can be written as \cite{Grice:1997, Branczyk:2010}
$\hat{\rho}(\omega_\text{i},\omega_\text{s}) = \underset{k}{\sum} \lambda_k \ket{\zeta_k^\text{i}} \bra{\xi_k^\text{s}}$,
with Schmidt coefficients $\lambda_k$ and orthonormal signal and idler modes with annihilation operators $\left\{\ket{\xi_k^\text{s}} \right\}$ and $\left\{\ket{\zeta_k^\text{i}}\right\}$, respectively. 
The lower the number of contributing mode pairs in $\hat{\rho}$, the higher is the state's purity
$\mathcal{P} = \underset{k}{\sum} \frac{1}{\lambda_k^2}$. 
For a photon counting detector with uniform spectral and temporal response all spectral-temporal information in the idler mode is lost and thus the spectral-temporal state of the heralded single photon (HSP) is obtained by tracing over the idler subsystem in $\hat{\rho}$, yielding the marginal signal state 
$\hat{\rho}_\text{s} = \mathrm{Tr}_\text{i} (\hat{\rho} (\omega_\text{i}, \omega_\text{s} ) )$.

\paragraph{Mode matching to the memory}
Once heralded, the signal photons are to be stored in a quantum memory, for which reason their spectral mode has to match the memory's spectral acceptance. 
For our discussion, we assume on-demand memories, whose storage and retrieval interaction is mediated by a control field \cite{Nunn:2007wj, Lukin:2003vn}. 
The control maps the incoming photon onto a stationary atomic spin-wave coherence $\hat{B}(\vec{x})$, which is distributed along the spatial dimensions $\vec{x}$ of the storage medium. 
Both, the spectral electric field amplitude $\hat{E}_\text{s}(\omega)$ and $\hat{B}(\vec{x})$ are linked by the memory kernel $K(z,\omega)$ as
$\hat{B}(z) = \underset{\omega_\text{s}}{\int} K(z,\tilde{\omega}) \hat{E}_{\text{s}} \left(\tilde{\omega}\right) \text{d} \tilde{\omega}$, 
with the control field spectrum defining the kernel's spectral characteristics. 
Therewith, we can define the memory efficiency as the expectation numbers of the resulting spin-wave excitations per incoming signal photon
%$$
\begin{equation}
\eta_\text{in} = \frac{\langle \hat{N}_B \rangle}{\langle \hat{N}_{E_\text{s}} \rangle}= 
\frac{\langle \hat{B}^\dagger \hat{B} \rangle}{\langle \hat{E_\text{s}}^\dagger \hat{E_\text{s}} \rangle} = \frac{\underset{z}{\int} |B(\tilde{z})|^2 d \tilde{z}}{\underset{\omega}{\int} |E_\text{s}(\tilde{\omega})|^2  \text{d}\tilde{\omega}}.
\end{equation}
%$$
Memory retrieval is the reverse process, $\hat{E}_\text{ret}(\omega_\text{s})=\underset{z}{\int} K^\text{r}(\omega_\text{s},\tilde{z}) \hat{B}(\tilde{z}) \text{d} \tilde{z}$, induced by a subsequent control pulse via the retrieval kernel 
$K^\text{r}(\omega_\text{s}, z)$. Similarly, the retrieval efficiency is 
$\eta_\text{ret} =\frac{\langle \hat{N}_{E_\text{ret}} \rangle}{\langle \hat{N}_B \rangle}$, yielding a total memory storage efficiency of $\eta_\text{mem} = \eta_\text{in}\cdot\eta_\text{ret}$. 
Here, we assume the memory acceptance mode, determined by $K(\omega,z)$, can be adjusted at will by appropriately shaping the control's spectral amplitude $E_\text{c}(\omega)$.
When considering our Raman memory system, which is operated in the low control energy regime \cite{Nunn:2007wj}, the memory kernel spectrum can be approximated by the Rabi-frequency 
$\tilde{\Omega}(\omega) \sim E_\text{c}(\omega)$, and thus corresponds directly to $E_\text{c}(\omega)$. 

Accordingly, source - memory interface optimisation requires maximisation of the excited spin-wave fraction. 
This can, for instance, be achieved when producing HSPs in a single mode ($\mathcal{P}\rightarrow 1$) that spectrally matches the memory kernel, which 
corresponds to the well-known single photon source design objective of generating pure states. 
For the case of a travelling-wave, GHz-bandwidth source, where the pump map dominates the JSA, the available design parameters are the pump and filter bandwidths. 
Considering solely the source, optimisation would require narrowband filtering to MHz, effectively reducing the filtered JSA to a horizontal stripe in $\omega_\text{s}-\omega_\text{i}$ space. 
However, for operation with a memory we also have to take temporal aspects into account, which have so far been ignored. 

\paragraph{Timing of single photon production}
A combined memory-source system will, in practice, be operated with a repetition rate $f_\text{rep} = \frac{1}{T}$, since read-in or retrieval events have to occur within a time bin $T$. 
Experimentally, $f_\text{rep}$ can, for example, be set by the pump laser of the SPDC process.  
For the read-in process this time bin size effectively limits the signal and idler duration (see Figure~\ref{fig_1} for an illustration), 
which can be understood intuitively, when considering on demand memories: 
Here control pulse generation is triggered by idler detection events, as idler detection also heralds the presence of a single photon to be inserted into the memory. 
If idler photons overlapped with adjacent time bins and were detected in one of these neighbouring bins, the generated control pulses would fall into these adjacent time bins as well. 
Accordingly, the control would not be synchronised with the SPDC signal photons, going into the memory. 
Without the presence of a control pulse in the time bin of the incoming single photon, it would not be stored and the observable storage efficiency would degrade. 
Equally, the SPDC signal photons should also not overlap with any other time bins than the one they are generated in, which corresponds to the time bin of the SPDC pump pulse. 
Otherwise, the heralded single photons would have a finite probability to fall into another time bin than the memory control pulse, leading to a similar reduction in storage efficiency. 
Hence we need to incorporate such time gating when optimising the pump and idler filter bandwidth parameters for maximum read-in efficiency. 

Mathematically, both time bin constraints are easily expressible when considering the joint temporal amplitude (JTA), i.e. the filtered JSA in the time domain. 
Here we can directly apply the required time bin restrictions by multiplying the 
JSA with gating functions $G(t_\text{i,s}, T)$ for the signal and idler photons. 
Such time gates can, for instance, be modelled by Heaviside functions with $G_\text{i,s}(t_\text{i,s}) = \text{rect}(t_\text{i,s} - \frac{T}{2})$. 
These are temporally centred on an SPDC pump pulse ($t=0$), whose duration, in turn, defines the time bin size $T$. 
To arrive at an expression for the memory read-in efficiency, we start with a train of SPDC pump pulses 
%$$
\begin{equation}
\Omega_\text{tot} (t,T)= \sum_{j=-\infty}^{+\infty} \Omega(t-j \cdot T).
\end{equation}
%$$ 
Without any filtering, the JSA $\tilde{f}(\omega_\text{i},\omega_\text{s})=\tilde{\Omega}_\text{tot}(\omega_\text{i},\omega_\text{s})$ is determined by the Fourier transform of the pump pulse train $\Omega_\text{tot} (t,T)$. 
Its frequency comb structure means, the JSA consists of small stripes for each tooth in the comb, oriented along a $45^\circ$-angle in $\omega_\text{s}$-$\omega_\text{i}$ space. 
Applying the idler filter results in the modified JSA $\widetilde{f}_\mathcal{T}(\omega_\mathrm{s},\omega_\mathrm{i})$.  
To return to the time domain, we Fourier transform the filtered JSA over both frequency dimensions and use the shift theorem to arrive at 
%$$
\begin{equation}
{f}_\text{F}(t_\text{t},t_\text{s}) = {\mathcal{T}}(t_\text{i}-t_\text{s}) {\Omega}_\text{tot}(t_\text{s}). 
\end{equation}
%$$
$ {\mathcal{T}}(t_\text{i}-t_\text{s})$ denotes the idler filter's amplitude transmission in the time domain, 
which restricts the temporal separation between signal and idler photons. 
We can now apply the time gates by straight forward multiplication with the JTA 
%$$
\begin{equation}
{f}_\text{G,F}(t_\text{t},t_\text{s}) = %G(t_\text{i}) G(t_\text{s}) {f}_\text{F}(t_\text{t},t_\text{s}) = 
G(t_\text{i}) G(t_\text{s})  {\mathcal{T}}(t_\text{i}-t_\text{s}) {\Omega}_\text{tot}(t_\text{s})
\end{equation}
%$$
and arrive at a marginal density matrix for the heralded single photons of 
%$$
\begin{equation}
\rho_\text{s}(t_\text{s}, t_\text{s}') = \int  {f}_\text{G,F} (t_\text{s}, t_\text{s}) {f}^{*}_\text{G,F} 
(t_\text{s}, t_\text{s})  \text{d}t_\text{i} \ket{1_\text{s} } \bra{1_\text{s}},
\end{equation}
%$$
where, for simplicity, higher order terms have been ignored \cite{Branczyk:2010}. 
Therewith, the efficiency of the memory is just given by 
\begin{equation}
\eta_\text{in} = \bra{K(t_\text{s},z)} \rho_\text{s} \ket{K(t_\text{s},z)}. 
\label{eq_2}
\end{equation}
The memory's input mode, represented by the memory kernel $K(t_\text{s},z)$, is now operating in the time domain and determined by the temporal pulse shape of the control.

\paragraph{Source design}
To summarise, the key parameters for the photon source are the SPDC pump pulse duration $T$, its repetition rate $f_\text{rep}$ and the idler filter bandwidth $\gamma$. 
The pump's pulse duration determines the spectral bandwidth of the unfiltered JSA, and thus the spectral correlations between signal and idler photons. 
The pump's effective time bin size $T$, in turn, defines the time bin sizes for storage and retrieval, since the idler photon detection triggers the memory control pulse \cite{Michelberger:2014} (Figure~\ref{fig_1}). 
Of course, if we aim to run the joint source - memory system with the highest possible clock rate, the pump's pulse duration will put an effective upper bound on the system's repetition rate with $f_\text{rep} \sim \frac{1}{T}$. 
The idler filter bandwidth $\gamma$ needs to be selected appropriately to produce heralded single photons that spectrally match the memory control pulses. 
Moreover, the temporal durations of the filtered idler and the heralded signal photons need to be confined within the memory time bin size $T$, set by the pump's repetition rate. 

We now investigate these parameters with the aim to optimise the source - memory interface, such that a produced heralded single photons can be stored with a high probability in $\eta_\text{in}$.  
To this end, we assume a train of Gaussian-shaped pulses, with electric field amplitudes 
\begin{equation}
\Omega_j (t) = \Omega_0 \cdot \exp{\left\{ - \frac{ \left( t - j \cdot T \right)^2 }{\sigma_\text{p}^2 }\right\}},
\end{equation}
is pumping the SPDC process. 
We also model the idler filter's electric field transmission by a Gaussian-shaped filter line, with a time domain representation of 
\begin{equation}
{\mathcal{T}}(t) = \exp{\left\{ - \left( \gamma \cdot t \right)^2 \right\}}. 
\label{eq_Gauss_filter}
\end{equation}
This matches the situation in the experiment (described below). 
Furthermore we express the parameters $T$ and $\gamma$ in units of $\sigma_\text{p}$. 
Assuming no constraints on our ability to generate specific memory control pulse configurations, 
we choose electric field amplitudes for the memory control that match the fundamental mode $\ket{\xi_1^\text{s}}$ of the SPDC signal photon. 
Therewith, we effectively apply mode filtering \cite{Brecht:2011} and select the signal and idler mode pair with the largest Schmidt coefficient $\lambda_1$. 
Picking the first Schmidt mode pair maximises the memory efficiency $\eta_\text{in} \sim  |\lambda_1|^2$, compared to any mode selection including higher order modes \cite{Nunn:2007wj}. 
With these assumptions we are now in a position to evaluate the influence of the remaining two design parameters ($T$ and $\gamma$). 

Figure~\ref{fig_2}~(c) illustrates the memory efficiency dependence on time bin size $T$ and idler filter bandwidth $\gamma$. 
For larger pump pulse separations $T$, i.e. long time bins, tight idler filtering yields better memory efficiency ($\gamma \rightarrow 0.2$). This result is expected in a situation when temporal overlap of the signal and idler pulse envelopes with adjacent time bins is negligible: 
Narrowing down the spectral bandwidth of the idler filter results in the production of heralded single photons in purer states \cite{Branczyk:2010}. 
Accordingly, the marginal signal photon spectrum contains fewer higher order Schmidt modes, and the mode filtering storage process results in lower losses, increasing the storage probability. 

This picture changes when the repetition rate goes up and the time bins become shorter. 
Now, the temporal broadening of narrow-band filtering the SPDC idler photons can cause significant temporal overlap with the pre- and succeeding time bins. 
The resulting probability for timing mismatch between memory control and heralded single photon input reduces the storage efficiency. 
Since this mismatch increases for increasingly narrower filters and shorter time bins, the optimal storage efficiency shifts towards looser idler filtering. 

In designing a source there is thus a trade-off between, on the one hand, the purity of the produced single photons and the associated mode selection probability by the memory control (via the herald filter bandwidth $\gamma$), and, on the other hand, the maximum repetition rate the joint source-memory system can operate at (via the pump time bin parameter $T$). 
The higher the latter is sought to be, the lower the former needs to be. 

As mentioned initially, high rates are, for instance, relevant for temporal multiplexing applications \cite{Nunn:2013ab}, where many single photon production as well as storage and retrieval attempts are required, and the photon source is run at rates on the order of the inverse pump pulse duration ($T \rightarrow \sigma_\text{p}$). 
In a realistic scenario, one would aim to operate the joint system with $T \gtrsim 2 \sigma_\text{p}$, as otherwise the pump pulses start to overlap significantly and one moves into a pseudo cw-regime. 
As our results in Figure~\ref{fig_2}~(c) show, at this, de facto, limiting time bin size of $T\sim 2\sigma_\text{p}$, the optimal idler filter bandwidth approaches $\gamma_\text{opt} \sim 0.9 \sigma_\text{p}$, which is significantly broader than the narrowband solution of $\gamma \sim 0.2$, obtained without any time binning constraints. 

Using these design considerations, we next look at the implementation of an experimental prototype for such a broadband, memory-interfacing source.  
Recently, we have used the resulting system to store GHz-bandwidth heralded single photons in a high time-bandwidth-product Raman memory \cite{Michelberger:2014}.

\section{Matching the source to the Raman memory\label{experiment}}

We briefly describe our targeted Raman memory system before we outline the experimental apparatus  
and present the obtained photon counting statistics. 
We then investigate the matching of the produced heralded single photons to the Raman memory's signal input channel. To this end, we characterise the signal photon heralding efficiency ($\eta_\text{her}$), i.e. the preparation probability of a single photon in the memory's input channel conditional on the detection of an idler photon, and the marginal spectrum of the SPDC signal photons. 
As their GHz-bandwidth is too narrowband for standard spectrometers, but too broadband for direct detection on a photodiode, we employ a technique based on sweeping their spectrum over a known filter. 

\begin{figure*}
\centering
\includegraphics[width=0.99\textwidth]{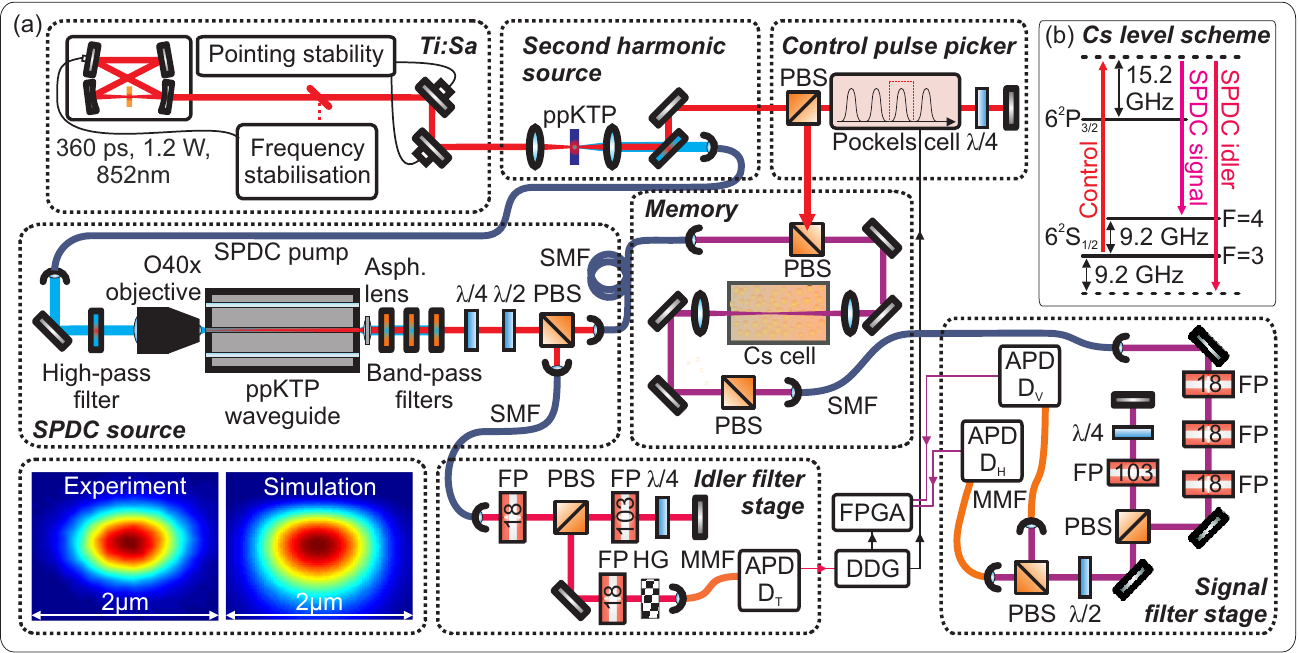}
\caption{
(a) Experimental set-up: see main text for details. 
(b) $\Lambda$-level system for the Raman memory in atomic Cs vapour \cite{Nunn:2007wj}
and optical fields for on-demand heralded single photon storage in the memory. 
For the two-photon Raman transition, the control couples to the $6^2\text{S}_{\frac{1}{2}}$ F$=3$  state and the heralded SPDC signal photon to the $6^2\text{S}_{\frac{1}{2}}$ F$=4$ state. 
The corresponding SPDC idler photon must be projected to the blue of the control by the $9.2\,\text{GHz}$ Cs hyperfine ground state splitting. 
}
\label{fig_3}      
\end{figure*}

\paragraph{Raman memory characterisitcs} 
We implement the Raman memory \cite{Nunn:2007wj, Reim2010, Reim2012, England2012, Michelberger:2014} in $70^\circ \text{C}$-warm atomic caesium (Cs) vapour. 
Operating on the D$_2$-line, with $852\,\text{nm}$ central wavelength, the memory relies on the $\Lambda$-system illustrated in Figure~\ref{fig_3}~(b). 
Signal photons are stored by transferring atomic population from the initial $6^2 S_{\frac{1}{2}}, \text{F} = 4$ state (memory empty) to the $6^2 S_{\frac{1}{2}}, \text{F} = 3$ state (memory charged). 
These two hyperfine ground states are separated by $\delta_\text{gs} = 9.2\,\text{GHz}$, which forms the upper limit on the bandwidth of storable heralded single photons. 
The population transfer is mediated by a strong control field, detuned by $\Delta_\text{m} = 15.2\,\text{GHz}$ from the $6^2 P_{\frac{3}{2}}$ excited state manifold. 
Control pulses of 
$\mathcal{E}_\text{c} \sim 15\,\text{nJ}$ pulse energy and 
$\Delta \nu_\text{c}\sim 1 \,\text{GHz}$ 
full-width-half-maximum (FWHM) spectral bandwidth are produced with a titanium-sapphire (Ti:Sa) laser. 
At this pulse energy, our memory operates in the low power regime, so the Raman memory's spectral acceptance bandwidth is set by the control pulse \cite{Nunn:2007wj}.

\paragraph{Experimental setup}
We use an $f_\text{rep} = 80\,\text{MHz}$ repetition rate Ti:Sa master oscillator to generate the SPDC pump and the memory control. 
Its pulse duration, on the order of $1\,\text{ns}$, and pulse time bins size of $12.5\,\text{ns}$ would position us at $T\sim 11$ in Figure~\ref{fig_2}~(c). 
While this would leave us with an optimal idler filter bandwidth on the order of $\gamma \sim 0.2 \sigma_\text{p}$, we aim to simulate the broadband case. 
Hence, for this proof-of-principle demonstration, we use a filter bandwidth of 
$\gamma \sim 0.85 \sigma_\text{p}$ which is close to $\gamma_\text{opt}$ \footnote{ 
We use an $80\,\text{MHz}$ laser system for reasons of availability. The draw-back of not having a GHz repetition rate laser system available is that the resulting memory efficiency of the interfaced source - memory system is less affected by projections into incorrect time bins.  
}. 

In the experiment, sketched in Figure~\ref{fig_3}~(a), the Ti:Sa output is frequency doubled with low efficiency ($\sim 9\,\%$) in a periodically-poled ppKTP crystal. The resulting SPDC pump pulses at $426\,\text{nm}$ wavelength (UV) have a FWHM spectral bandwidth of 
%$\Delta \nu_\text{p} \sim 1.2 \,\text{GHz}$  % Sech pulses 1.19
$\Delta \nu_\text{p} \sim 1.3 \,\text{GHz}$  % Gaussian pulses 1.29
and are guided to the photon source, with the unconverted fundamental at $852\,\text{nm}$ wavelength (IR) separated on a low pass filter and sent into a Pockels cell for control pulse picking from the Ti:Sa pulse train \cite{Michelberger:2014}.   
The SPDC source is implemented in a $2\,\text{cm}$ long nonlinear ppKTP waveguide (\textit{AdvR}), heated to $T_\text{KTP} \sim 40^\circ \text{C}$, using a $3\,\mu\text{m}$ wide and $6\,\mu\text{m}$ deep (FWHM) waveguide channel. 
A microscope objective (\textit{Olympus 40X}) couples the UV pump into the fundamental transverse mode \cite{Karpinski:2009}, which have a total transmission efficiency of $\sim 10\,\%$ through the waveguide chip. 
The type-II SPDC emits signal and idler photons, at $852\,\text{nm}$ central wavelength, into the fundamental transverse mode \cite{Karpinski:2012} (see Figure~\ref{fig_3}~(a) for mode images). 
These are separated from the UV on another low pass filter and fluorescence noise is filtered by a series of $852\pm2.5\text{nm}$ bandpass filters. 
Signal and idler are split on a polarising beam splitter (PBS). 
The idler is filtered and detected. 
Meanwhile, the signal is delayed in $91\,\text{m}$ of single mode fibre (SMF), to allow sufficient time to pick the memory control pulses, triggered by idler detection events (feed forward operation \cite{Michelberger:2014,Sinclair:2014}). 

The idler filter contains a holographic grating (HGF) with a $100\,\text{GHz}$-wide reflection band and a series of 4 air-spaced Fabry-Perot (FP) filters; two etalons with a finesse of $\mathcal{F} = 12$ and $18.2 \,\text{GHz}$ free-spectral range (FSR) and one, double passed, etalon with $\mathcal{F} = 68.9$ and $\text{FSR}=103\,\text{GHz}$. 
While the HFG and the $\text{FSR} =103\,\text{GHz}$ etalons suffice to select the correct frequency mode, the additional $\text{FSR} =18\,\text{GHz}$ etalons suppress single photon fluorescence in the wings of the $\text{FSR}=103\,\text{GHz}$ etalon filter lines. 
The total transmission profile has a
$\text{FWHM} = 1.4\,\text{GHz}$  % Gaussian pulse model
%$\text{FWHM} = 0.94\,\text{GHz}$  % Sech pulse model
%$\text{FWHM} = 1.4\,\text{GHz}$ %no idea what that number is
and is well described by the Gaussian line $\widetilde{\mathcal{T}}(\nu_\mathrm{i})$ (Fourier transform of Equation~\ref{eq_Gauss_filter}). 
Having the Ti:Sa laser's frequency ($\nu_\text{Ti:Sa}$) set to the memory control channel in the $\Lambda$-system (Figure~\ref{fig_3}~(b)), the SPDC signal photons need to be projected by $9.2\,\text{GHz}$ towards the red to achieve two-photon resonance for Raman storage \cite{Nunn:2007wj}. 
To this end, we adjust the centre of the idler filter's transmission line to $\nu_\text{i}^\text{f} = \nu_\text{Ti:Sa} + 9.2\,\text{GHz}$, yielding heralded single photons with $\nu_\text{s} = \nu_\text{Ti:Sa} - 9.2\,\text{GHz}$ central frequency. 
Thanks to the broad phase-matching bandwidth of $\sim 100\,\text{GHz}$ (FWHM), this is possible without any significant reduction in the SPDC pair production rates. 
Post filtering, idler photons are detected on a single photon avalanche photodiode (APD, \textit{Perkin Elmer}). 
Its output signals are sent into a digital delay generator (DDG), which, firstly, feeds into a field programmable gate array (FPGA) for photon counting, and, secondly, into the Pockels cell unit. 
After appropriately delaying the idler detection signals with the DDG, the Pockels cell can use them as a trigger to pick control pulses from the unconverted Ti:Sa pulses, such that arrival times at the memory are synchronised with the heralded single photons. 

For storage in the Cs, the heralded SPDC photons also feed into the vapour cell. 
Any photons, transmitted through the memory (i.e. not stored during read-in), or retrieved from the memory, are subsequently separated from the orthogonally polarised memory control pulses and filtered by a second filter stage. 
It comprises three etalons with $\text{FSR} =18$ GHz and another, double-passed $\text{FSR}=103$ GHz etalon. 
These suppress control pulse leakage and noise from the memory storage medium \cite{Michelberger:2014, Reim:2011ys}. 
The resulting spectral filter transmission line $\widetilde{\mathcal{T}}(\nu_\mathrm{s})$ is also well approximated by a Gaussian profile (Equation~\ref{eq_Gauss_filter}) with a FWHM filter bandwidth of 
$\Delta \nu_\text{s} \sim 1.1\,\text{GHz}$		% Gauss pulse model
%$\Delta \nu_\text{s} \sim 1.06 \,\text{GHz}$ 	% Sech pulse model
%$\Delta \nu_\text{s} \sim 1\,\text{GHz}$ 		% no idea where this number comes from
and the filter resonance positioned at the SPDC signal frequency (i.e. $\nu_\text{s}^\text{f} = \nu_\text{s}$). 
Post filtering the signal is sent into a Hanbury-Brown-Twiss (HBT) type setup \cite{HBT:1956}, where it split 50:50 and detected on two multi-mode-fibre (MMF) coupled APDs, whose outputs also feed into the FPGA. 
By counting coincidences $c_{\text{H}|\text{T}}$ and $c_{\text{V}|\text{T}}$ between the detection events $c_\text{T}$, registered by the idler detector, and either signal detector ($c_\text{H}$, $c_\text{V}$), we can measure the produced heralded single photons in the input and retrieval time bin of the memory. 
Recording triple coincidences $c_{(\text{H} \& \text{V})|\text{T}}$ between all three detectors furthermore allows us to observe the photon statistics of the signal through the $g^{(2)}$ autocorrelation function \cite{Loudon:2004gd, Michelberger:2014}. 

We now discuss the characterisation of the generated heralded single photons. These characterisation measurements do not require single photon storage, so no control pulses are applied to read the photons into and out of the Raman memory.

\begin{figure*}
\includegraphics[width= 0.98 \textwidth]{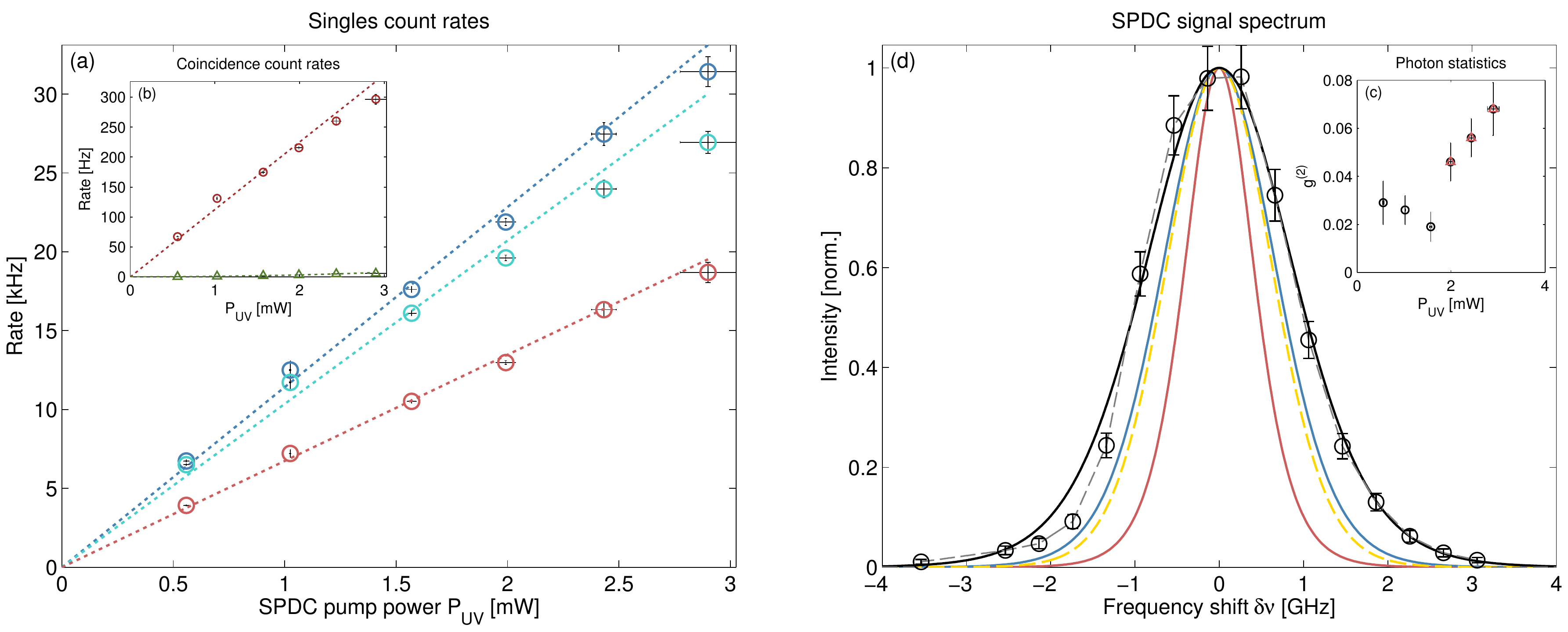}
\caption{
(a): Count rates of SPDC signal and idler photons. Red and blue data illustrate the signal and idler rates $c_\text{T}$ and $c_\text{H}+c_\text{V}$, respectively. 
The cyan data show the idler rates after temporal filtering by the digital delay generator (Figure~\ref{fig_2}~(a)). Dotted lines are a linear fits in $P_\text{UV}$. 
(b): Conditional detection rates $c_{\text{s}|\text{T}}$ of SPDC signal photons, registered on APDs D$_\text{H}$ and D$_\text{V}$, given an idler event on APD D$_\text{T}$. Red and green data are coincidences for signal and idler photons emitted within the same and in different UV pump pulses, respectively. Dotted lines are linear fits in $P_\text{UV}$. 
(c): $g^{(2)}$ autocorrelation for the produced heralded single photons. 
(d): Heralded single photon spectrum $S_\text{HSP}(\nu)$ (blue line), obtained from the convolution between $S_\text{HSP}(\nu)$ and the signal filter transmission line $\widetilde{\mathcal{T}}(\nu)$, centred on the Raman memory's input signal channel (i.e. $\delta\nu = \nu - \nu_\text{s}^\text{f}$). 
The black line shows the fitted convolution onto the measurement points (black points), which are interpolated by a grey dotted line. For comparison, the yellow line displays the marginal SPDC spectrum, expected by the JSA (Figure~\ref{fig_2}~(c)), and the red line denotes the memory control pulse spectrum. 
}
\label{fig_4}      
\end{figure*}

\paragraph{Photon production rates} 

Figure~\ref{fig_4}~(a) presents the photon count rates, registered by each detector (APD$_\text{T}$, APD$_\text{H}$, APD$_\text{V}$) in the setup \footnote{Due to the heavy spectral filtering of the SPDC output prior to detection (Figure~\ref{fig_3} (a)), the photon count rates are lower than one would expect from an unfiltered source. Filtering also makes memory and detector noise the dominant noise sources.}.
While idler detection events could be fed-forward directly to trigger the control pulse preparation, the temporal filtering by the DDG suppresses false triggers by detector dark counts or uncorrelated single photon fluorescence noise, falling into the selected spectral mode of the idler. 
Imposing a fixed minimum delay between successive input events, the DDG slightly reduces the rate $c_\text{T}$ of the transmitted signals, resulting in a maximum repetition rate of $f_\text{rep}= c_\text{T} \lesssim 25 \,\text{kHz}$ for the available UV pump power range ($P_\text{UV} \lesssim 3\,\text{mW}$) \footnote{
While this rate is lower than $f_\text{rep}^\text{max} \sim \frac{1}{\Delta \nu_\text{P}} \sim 1\,\text{ns}$, required to fully exploit the memory bandwidth, our Pockels cell currently imposes a technical upper limit of $f_\text{rep} \lesssim 12\,\text{kHz}$ \cite{Reim2010,Michelberger:2014}. 
For our proof-of-principle demonstrations the idler detection rates are thus sufficient. 
}. 

The SPDC process is operated in the low pump power regime \cite{Eisenberg:2004, Krischek:2010ys}. 
Consequently, the coincidence events between the idler trigger and each of the two signal detectors, 
$c_{\text{s}|\text{T}} = c_{\text{H}|\text{T}} + c_{\text{V}|\text{T}}$, 
displayed in Figure~\ref{fig_4}~(b), increases, to good approximation, linearly with the UV pump power ($P_{UV}$). 
We can estimate the contribution from uncorrelated signal photons therein 
by counting the coincidences between idler and signal detectors for a pre- or succeeding time bin of the $80\,\text{MHz}$ SPDC pump pulse train \cite{Ramelow:2013}. 
Adding a $12.5\,\text{ns}$ time delay between the FPGA channel for APD$_\text{T}$ and the channels for both signal APD$_\text{H}$ and APD$_\text{V}$, we obtain the rates $\bar{c}_{s|T}$ in Figure~\ref{fig_2}~(c). These amount to $< 2 \,\%$ of the total coincidences $c_{s|T}$ (at $P_\text{UV} \sim 3\,\text{mW}$),  
and are subtracted from $c_{s|T}$ in the following.

\paragraph{Heralding efficiency}
With these background subtracted counts, we can now immediately determine the source heralding efficiency 
\begin{equation}
\eta_\text{her} = \frac{(c_{s|T} - \bar{c}_{s|T})}{c_T \cdot T_\text{s} \cdot \eta_\text{det}} \approx 25 \pm 3 \,\%. 
\label{eq_hereff}
\end{equation}
to express the probability of having a single photon present at the memory input upon every storage attempt, i.e. whenever a control pulse is triggered.  
Since we want to evaluate the actual single photons arriving at the memory, we account for the 
optical propagation losses between the memory and the signal detectors, $T_\text{s} \approx 10 \pm 1\,\%$, and the APD detection efficiencies, assumed as $\eta_\text{det} \sim 50\,\%$. 
On average, we send a single photon into the memory on every fourth trial. 
We estimate $\eta_\text{her}$ to be limited by optical losses in the beam path of the heralded single photons to the memory (coupling to and propagation in SMF) and transmission losses in the waveguide.

\paragraph{Photon number statistics}
Next we determine how close the source output is to a true single photon. 
Using the HBT-type signal detection setup in Figure~\ref{fig_3}~(a), we measure the heralded second-order autocorrelation function, defined as \cite{Loudon:2004gd}  
\begin{equation}
g^{(2)} = \frac{c_{(H\&V)|T}\cdot c_T}{c_{H|T} \cdot c_{V|T}}. 
\end{equation}
For a true single photons in the signal arm, $g^{(2)}$ should approach 0, while for a single-mode coherent state or thermal noise it would reach values of 1 or 2, respectively. 
Using the coincidences of each signal APD with the temporally filtered idler events 
($c_{\text{H}|\text{T}}$, $c_{\text{V}|\text{T}}$), as well as the triple coincidences between all three detectors ($c_{(\text{H}\&\text{V})|\text{T}}$), we consistently achieve $g^{(2)}\lesssim 0.08$ for all available UV pump power values (Figure~\ref{fig_4}~(c)). 
Higher order photon pair emissions are thus negligible and we indeed insert at most one single photon into the memory upon every storage trial.

\paragraph{Spectral matching of heralded single photons}
We can now analyse, how well these single photons actually interface with the Raman memory's input channel. 
To this end, we need to determine the heralded single photon spectrum. 
A direct measurement is difficult, since it would, e.g., require a single-photon spectrometer with sub-GHz resolution. 
However, a bandwidth measurement is nevertheless possible by 
sweeping the central frequency (${\nu}_{\text{s},0}$) of the single photon spectrum over the signal filter, whose resonance is kept constant at $15.2\,\text{GHz}$ blue-detuning of the $6^2 \text{S}_\frac{1}{2}, \text{F} = 4 \rightarrow 6^2 \text{P}_\frac{3}{2}$ transition in the Raman memory's $\Lambda$-system (Figure~\ref{fig_3}~(b)). 
We do this by modifying the central frequency $\nu_\text{Ti:Sa}$ of the Ti:Sa laser. 
Because of energy conservation, any change in $\nu_\text{Ti:Sa}$ by $\delta \nu$ 
shifts the SPDC pump pulse frequency ($\nu_\text{UV}$) by $2\cdot \delta \nu$ and, in turn, the SPDC signal and idler photons by $\delta \nu$.
Scanning the Ti:Sa frequency over the signal filter resonance $\widetilde{\mathcal{T}}(\delta {\nu} = \nu-\nu^\text{f}_\mathrm{s})$ and measuring the transmitted coincidences yields a count rate level proportional to  
\begin{equation}
\tilde{c}_{s|T}(\delta {\nu}) = \frac{c_{s|T}({\delta \nu}) }{c_{s|T}(0)} = \int_{\nu} 
\widetilde{\mathcal{T}}(\delta {\nu}) \cdot 
S_\text{HSP}\left( \delta {\nu} - {\nu'} \right) \cdot \text{d} {\nu'}, 
\label{eq_HSPconv}
\end{equation}
the convolution between the single photon spectrum and $\widetilde{\mathcal{T}}(\delta {\nu})$, whereby we already normalise the count rates. 
Notably, thanks to the broad emission bandwidth of the unfiltered SPDC process (Figure~\ref{fig_2}~(b)), 
the idler filter resonance can be kept constant without affecting $c_{\text{s}|\text{T}}(\delta \nu)$, when varying $\delta \nu$. 
This considerably simplifies the experimental procedure. 
To extract the single photon bandwidth from Equation~\ref{eq_HSPconv}, we assume again a Gaussian pulse spectrum, given by
\begin{equation}
S_\text{HSP}(\nu) \sim \exp{\left\{ -4 \pi^2 \left( \Delta t \right)^2 \cdot \left( \nu - \nu_0 \right)^2 \right\}}. % Gaussian pulse model
% \text{sech}^2(\pi^2 \cdot \Delta t \cdot \nu). % Sech pulse spectrum 
\end{equation}
We aim to extract the parameter $\Delta t$ to give the FWHM spectral bandwidth of the heralded single photons 
${
\Delta \nu_\text{s} = \frac{\sqrt{\ln{\left( 2 \right)}}}{\pi \cdot \Delta t}. 
}$
%${
%\Delta \nu_\text{s} = \frac{2\cdot \text{arcsech}(1/\sqrt{2})}{\pi^2 \cdot \Delta t} % sech pulses
%}$. 
For this purpose, we optimise $\Delta t$ to obtain the closest fit between the measured coincidence rates $\tilde{c}_{s|T}(\delta \nu)$ and the convolution of $S_\text{HSP}(\nu)$ with the known filter function $\widetilde{\mathcal{T}}(\delta \nu)$ (Fourier transform of Equation~\ref{eq_Gauss_filter}) \footnote{
	While for Gaussian pulses, we could rely on the addition of variances, our presented method is more 
	generally and can be applied to other pulse shapes (e.g. sech-pulses). 
}. 

Figure~\ref{fig_4}~(d) compares the optimised convolution with the measured data $\tilde{c}_{s|T}(\delta \nu)$ and shows the resulting single photon spectrum. 
We obtain a single photon spectral bandwidth of 
%$\Delta \nu_\text{s} =  1.69 \pm 0.06 \,\text{GHz}$, 	% Sech pulse model
$\Delta \nu_\text{s} =  1.78 \pm 0.06 \,\text{GHz}$, 	%	 Gaussian pulse model 
which we can contrast with the marginal SPDC signal spectrum, expected from the JSA in Figure~\ref{fig_2}~(a). 
For our signal and idler filter parameters, the latter predicts a bandwidth of  
$\Delta {\nu}_\text{s}^\text{JSA} =1.61 \,\text{GHz}$, 		% Gaussian pulse model
%$\Delta {\nu}_\text{s}^\text{JSA} =1.54 \,\text{GHz}$, 	% Sech pulse model
matching our experimental values closely. 
From the filtered JSA (Figure~\ref{fig_2}~(a)), we can also estimate the purity of the heralded single photons to $\mathcal{P} \sim 77 \,\%$. 
As desired by the idler filter choice, the heralded single photons fit into the time bin of the memory control pulse, for which reason their spectrum is broader than that of the control, with a 
$\text{FWHM}$ of $1.06\,\text{GHz}$ % Gaussian pulse model 
%$\text{FWHM}$ of $0.97\,\text{GHz}$ % Sech pulse model 
(also displayed Figure~\ref{fig_4}~(d) for comparison). 

Due to the sub-unity values of the purity $\mathcal{P}$ and the residual spectral mode mismatch between the single photons and the memory control, the memory efficiency is expected to reduce compared to a perfectly mode-matched input signal. 
Such a signal can, e.g., be generated by a low-intensity pick-off from the control pulse, and has been used in  our previous experiments \cite{Reim2010, Reim:2011ys, England2012}.  
Since the resulting coherent state \cite{Loudon:2004gd} input signal and the control derive from the same laser pulse, their spectral modes match.  
Using these as the memory's signal input, a storage efficiency of 
$\eta^\text{coh}_\text{in} \approx 51 \pm 2\,\%$ can be obtained \cite{Michelberger:2014}. 
Conversely, when sending the heralded single photons into the memory, the read-in efficiency is lowered by a factor $\sim 0.76 \pm 0.1$ to $\eta^\text{HSP}_\text{in} \approx 39 \pm 3\,\%$ \cite{Michelberger:2014}.

Given our earlier simulation results of Figure~\ref{fig_2} we would expect a reduction of this size:  
Since our laser system operates only at $80\,\text{MHz}$ repetition rate, our positioning along the  SPDC pump time bin size in Figure~\ref{fig_2}~(c) lies around $T \sim 11$, 
% for the $\sim 1.3\,\text{GHz}$ SPDC pump,
so our chosen filter bandwidth of $\gamma \sim 0.85$ should result in a reduction by a factor of $\sim 0.8$ over the mode-matched case. 
On the one hand, we are seeing the degradation in memory efficiency from choosing a broader idler filter, that filters the JSA less tight, reducing the purity of the heralded SPDC signal photons and, in turn, $\eta_\text{in}$. 
On the other hand, there are no effects from time-bin constraints on the storage efficiency yet, i.e. generating single photons with long pulse duration by narrowband idler filtering is not yet penalised. 
If we could increase the repetition rate of the system, the temporal binning effects would  become important, and the optimal filter parameter $\gamma$ would shift towards the value we have chosen in the experiment. 
The presented setup is thus only a prototype version, which uses the design parameters required for a  high repetition rate scenario, allowing us to estimate the achievable performance of an interfaced memory-source system at high rates. 
However, at present the implementation of such a reliable interface between a single photon source and a Raman memory is hindered by background noise issues on the memory-side \cite{Michelberger:2014, Saunders:2016, Nunn:2016,Kaczmarek:2017}, which have to be managed  first.

\section{Conclusion}

In this article, we discussed the design requirements for travelling-wave SPDC-based heralded single photon sources that are to be interfaced with on-demand quantum memories, and presented the characterisation of a prototype system. 
A joint source - memory system is an appealing solution to achieve close to deterministic operation with probabilistic heralded single photon sources via temporal multiplexing \cite{Nunn:2013ab}. 
To enable high single photon production rates, it is ideal to store broadband heralded single photons, at high repetition rates, in a memory with a large time-bandwidth product, allowing many storage and retrieval trials within the memory's lifetime. 
To this end, heralded photons should be spectrally pure, with bandwidths matching the spectral-temporal input channel of the memory, and temporally separable so they can be addressed by individual control pulses. 

Our analysis revealed an inherent trade-off between these objectives, when spectral engineering SPDC signal photons through spectral filtering of the heralding idler photon \cite{Grice:1997,  Branczyk:2010}. 
While for low repetition rate systems producing single photons with highest purity proves optimal, the associated tighter spectral filtering leads to false projections into neighbouring time bins at higher repetition rates. 
When seeking to operate the source at the highest possible repetition rates, a filter bandwidth on the order of the pump's time bin duration enables the largest memory read-in efficiencies for the produced heralded single photons, as it balanced the photons' temporal and spectral properties.

Based on this optimal filter solution, we have implemented a source prototype that we have recently interfaced with a Raman memory \cite{Michelberger:2014}. To characterise the quality of the obtainable interface, we demonstrated the desired temporal confinement of the produced heralded single photons to the memory's read-in time bin as well as their spectral matching to the memory's spectral acceptance profile \cite{Nunn:2007wj}. 
For this purpose, we presented a measurement for the spectral bandwidth of heralded single photons in the GHz-regime. 
In a final step we could show that the influence of single photon's spectral properties on their read-in efficiency into the Raman memory tied in with the expectations we obtained our initial design analysis.

\subsection{Acknowledgement}
We thank William S. Kolthammer, Duncan England and Marco Barbieri for helpful discussions and assistance with source optimisation. 
We acknowledge Nathan Langford for early contributions to the experimental design of the source. 
We thank Justin Spring for assistance with the FPGA and Benjamin Brecht for helpful discussions on the source design. 
This work was supported by the UK Engineering and Physical Sciences Research Council (EPSRC; EP/J000051/1 and Programme Grant EP/K034480/1), the Quantum Interfaces, Sensors, and Communication based on Entanglement Integrating Project (EU IP Q-ESSENCE; 248095), the Air Force Office of Scientific Research: European Office of Aerospace Research \& Development (AFOSR EOARD; FA8655-09-1-3020), EU IP SIQS (600645), the Royal Society (to J.N.) and EU ITN FASTQUAST (to P.S.M.).  I.A.W. acknowledges an ERC Advanced Grant (MOQUACINO).

%\bibliographystyle{naturemag}
%%\bibliographystyle{svjour.bst}
%\bibliography{References}

\end{document}